%% file: main.tex
\documentclass[sigconf,nonacm]{acmart}

\newcommand{\para}[1]{\smallskip\noindent {\bf #1}}

\long\def\comment#1{}
\newcommand{\ie}{{\em i.e.}}
\newcommand{\eg}{{\em e.g.}}
\newcommand{\sysname}{{\textbf{SAMS~}}}
\usepackage{xcolor}

\long\def\delete#1{}
\usepackage{multirow}
\begin{document}

\makeatletter
\def\@copyrightspace{\relax}
\makeatother

\title{Spatial Aware Multi-Task Learning Based Speech Separation}
\author{Wei Sun, Mei Wang, Lili Qiu}
\affiliation{
\institution{The University of Texas at Austin}
\country{USA}
}
\email{{weisun, mei, lili}@cs.utexas.edu}

\input{abstract}
\maketitle

\input{intro}

\input{related}
\input{approach}

\input{eval}
\input{conclusion}


\bibliographystyle{abbrv}
\bibliography{tracking}

\end{document}

%% file: abstract.tex
\begin{abstract}
During the Covid, online meetings have become an indispensable part of our lives. This trend is likely to continue due to their convenience and broad reach. However, background noise from other family members, roommates, office-mates not only degrades the voice quality but also raises serious privacy issues. In this paper, we develop a novel system, called Spatial Aware Multi-task learning-based Separation (SAMS), to extract audio signals from the target user during teleconferencing. Our solution consists of three novel components: (i)  generating fine-grained location embeddings from the user's voice and inaudible tracking sound, which contains the user's position and rich multipath information, (ii)   developing a source separation neural network using multi-task learning to jointly optimize source separation and location, 
and (iii) significantly speeding up inference to provide a real-time guarantee. Our testbed experiments demonstrate the effectiveness of our approach. 
\end{abstract}

%% file: intro.tex
\section{Introduction}
\label{sec:intro}

\para{Motivation:} Online meetings play an indispensable role in our daily life. During the Covid-19, it is the only means to connect for many people. Kids depend on it for education, adults rely on it for work, and friends count on it for socialization. Moreover, its importance will likely go well beyond Covid-19 as many companies will continue to allow work-from-home, and online classes will likely be the future trend due to convenience and broad reach.

Over 70\% US households have two or more people and the average household size across the world is 4. While one is participating in an online meeting or taking an online class, other house members may generate sound. Unlike videos, voice signals can travel across rooms and result in significant interference unintentionally. 
This both degrades the audio quality and raises serious privacy issues. 

\para{Existing work:} 
There has been significant work on signal source separation. Earlier works use signal processing, such as PCA and ICA. More recent works use machine learning (ML) to further improve the separation accuracy. Videos can also be used to improve source separation~\cite{audio-visual-sep,audio-visual-cocktail,SE-doppler} since the camera captures mouth position and movement. However, video requires good lighting conditions, has a limited field of view, and raises significant privacy concerns. \cite{internet-footprint} also reports turning off a camera during a teleconferencing reduces environmental footprint of a meeting by 96\%. 

Despite considerable works, existing works primarily focus on using raw audio samples for source separation. User location can have a significant impact on source separation but has not been explicitly considered until recently. Some recent works use the ground truth location for source separation and report significant benefits ~\cite{ multi1, multi2}. The existing video-audio work also assumes the mouth location is accurate, hence it cannot be directly applied to audio-only solutions since acoustic tracking has a larger error. CoS~\cite{jenrungrot2020cone} uses a binary search for the azimuth direction, which may fail due to multipath cancellation in a wider beam.



\para{Our approach:}  To preserve privacy and efficiency, we seek an audio-only solution to explicitly estimate the user's spatial information including position and multipath profiles for source separation. Since the spatial information contains non-negligible error, we need to explicitly consider the localization error in source separation. 

\begin{figure}[t]
  \centering
  \includegraphics[width=0.49\columnwidth]{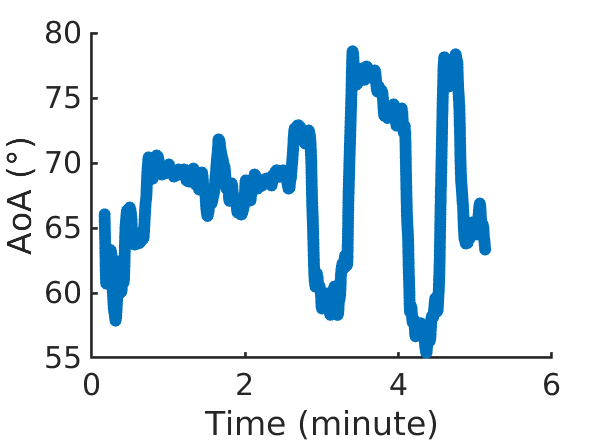}
  \includegraphics[width=0.49\columnwidth]{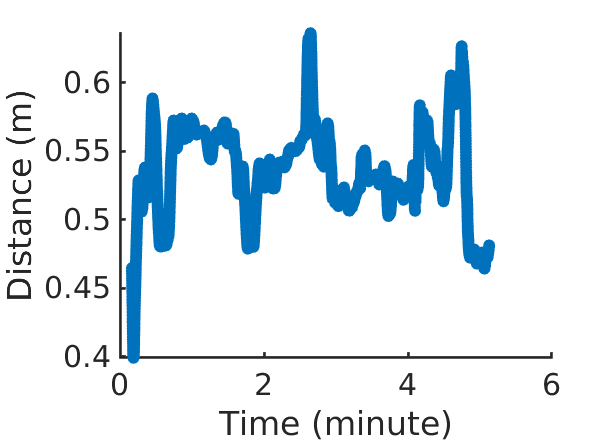}\hspace{-0.1in}
  \vspace*{-0.15in}
  \caption{\label{fig:motion_curve} Speaker movement during spontaneous speech (collected using depth camera).}
  \vspace*{-0.28in}
\end{figure}

Consider a user using a computer to join an online meeting and other people talking in the background. Since the user is close to the computer during the meeting, a small movement can result in a large difference in the Angle of Arrival (AoA). Our measurements in Figure~\ref{fig:motion_curve} show that there is frequent head movement that yields over 20 degree change in the AoA and 20 cm change in the distance when a user speaks spontaneously. Therefore it is useful to track the user's location and use it for source separation. 

\begin{figure}[t]
  \centering
  \includegraphics[width=\columnwidth]{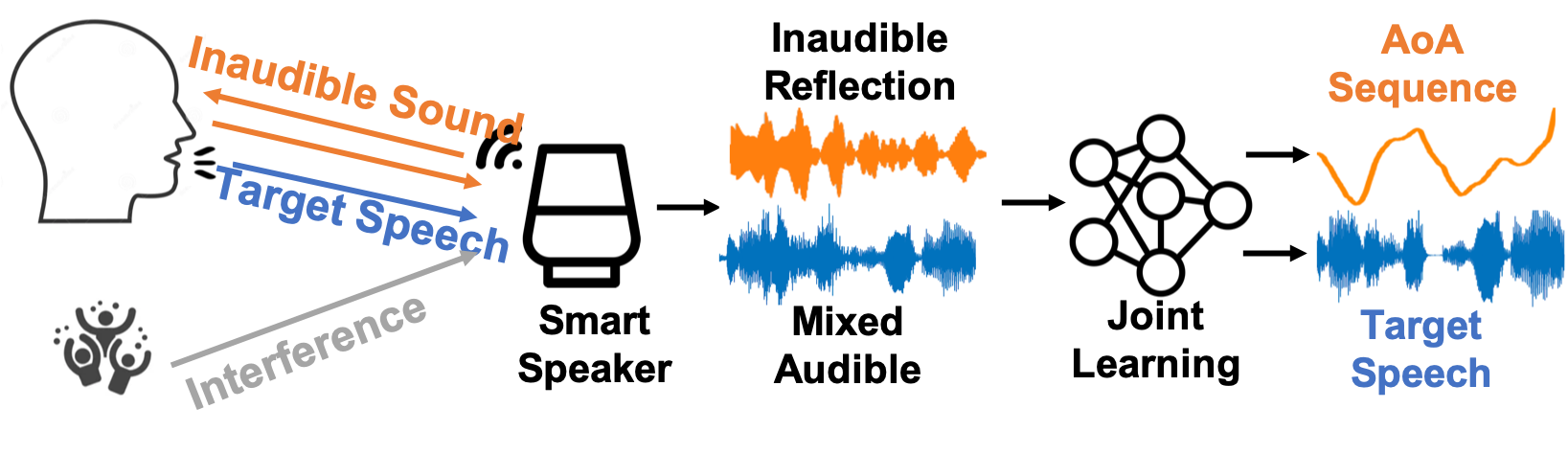}
  \vspace{-0.4in}
  \caption{\label{fig:system} \sysname automatically removes acoustic interference and ambient noise for online meetings.}
  \vspace{-0.3in}
\end{figure}

As shown in Figure~\ref{fig:system}, we develop a system that automatically removes the interference by explicitly estimating the user location and multipath, and using the estimated spatial embeddings to enhance source separation accuracy.


While there has been existing work on localizing a user using inaudible signals (\eg, \cite{CAT,apnea}) or audible signals (\eg, \cite{romit2020,MAVL}), our work is the first that leverages both audible and inaudible signals to achieve high localization accuracy. More specifically, we extract masks in the Time-Frequency (TF) domain from audible signals and use the mask to select TF bins dominated by the target user's voice to improve the localization accuracy of the voice signal under interference. Meanwhile, we let the computer generate inaudible acoustic chirps to track user position. We feed location profiles from both audible and inaudible signals to the 3D convolutional layer to generate spatial embeddings, which will serve as the input to source separation. These spatial embeddings contain not only the user's location but also rich multipath information, and play an important role in source separation.

We then leverage the user's location to improve the source separation by developing a novel multi-task learning framework to jointly learn the source separation and location. Instead of treating the estimated position as an input feature to the separator,  we explicitly take into account of the position estimation error in the multi-task loss function and guide the network to jointly learn the separation and localization by establishing the consistency between the target user speech and position.





In order to achieve real-time processing during an online meeting, we make following enhancements: (i) leveraging casual convolution (\ie, using only past samples) and non-casual convolution (\ie, using future samples) with a small look-ahead window to reduce response time, (ii) cache previous intermediate results in the neural network and reuse whenever possible, and (iii) further optimize the computation graph with Microsoft Onnxruntime~\cite{bai2019}.  According to ~\cite{realtime2}, 150 ms is a recommended one-way latency. \sysname processes audio every 90 ms within 42 ms on a laptop without GPU. So the total latency is 132 ms, well below 150 ms. 
We implement our approach on a laptop with dual speakers and a microphone array. We evaluate its performance with different environments, users, SNR, and interference sources. Our results show that \sysname achieves 10.71 - 13.61 dB Scale-invariant Signal-to-Noise Ratio (SiSNR) under a varying number of interfering users. This is 3.4-5.0 dB improvement over Conv-TasNet~\cite{Conv-TasNet}, 1.4-9.58 dB improvement over PHASEN~\cite{PHASEN}.

Our contribution can be summarized as follows:
(i) We generate a mask based on the target user's voice signals and apply it to localize the target user's voice under interference. We further estimate the 2D MUltiple SIgnal Classification (MUSIC) profile using inaudible signals and combine it with the audible location profiles using an LSTM network to generate spatial embeddings, which captures the acoustic multipath profiles. 
(ii)  We propose a multi-task learning framework to jointly learn source separation and location. We develop a novel pre-mask learning network that leverages the spatial embeddings from both audible and inaudible signals along with raw audible signals. By leveraging multi-task learning, it can simultaneously improve the localization and separation accuracy. (iii)  We implement our approach and significantly speed up the inference time to provide real-time guarantees. We demonstrate its significant performance benefits over existing approaches.  
This work has received {\em IRB approval}.

%% file: related.tex
\section{Related Work}
\label{sec:related}

Speech separation is a classic problem arising in many contexts.
We classify related work to single channel based separation, multi-channel based separation, and leveraging context information. 

\para{Single-channel based separation:} 
When the received signal has one channel, researchers exploit the inherent vocal features and speech content to extract the independent components from mixture signals. Most approaches ~\cite{wavenet, TCNN} use the spectrogram of the speech signal, computed by the short-time Fourier transform (STFT). A mask matrix can be estimated to filter out the target source from the mixture spectrogram.
Embeddings~\cite{deepcluster} are used to encode TF bins to higher dimensions for clustering. PHASEN~\cite{PHASEN} develops a two-branch learning framework to improve phase estimation. Meanwhile, time domain based encoders and decoders ~\cite{Conv-TasNet, dprnn} are proposed to replace STFT. SepFormer~\cite{sepformer} further integrates multi-head self-attention.
These approaches have shown good performance in the WSJ0MIX synthetic dataset. However, the single-channel input is limited to exploit crucial spatial information.

\para{Multiple-channel based separation:} Animals have developed multiple ears through millions of years of evolution. 
Similarly, more receiver channels can significantly improve separation performance in theory and practice. 
Independent component analysis (ICA)~\cite{saruwatari2003blind} separates the mixture into additive subcomponents by assuming non-Gaussion source signals and the number of sources to be smaller than the number of channels. Beamforming algorithms ~\cite{MVDR} leverage spatial information to strengthen the signal in target directions and null the interference from unwanted directions.
~\cite{multi1} proposes to iteratively run classic beamforming and separation to guide the network to focus on the appropriate direction. Recent work ~\cite{multi2, xu2021generalized} develops end-to-end learning of complex covariance matrix to predict spatial filter.
These works require the source position as an additional input for source separation.

\para{Leveraging context information:} 
Recent works start to exploit audio and visual information together~\cite{audio-visual-cocktail, VisualVoice}. VisualVoice \cite{VisualVoice} associates a speaker's facial appearance with source separation since it reveals gender, age and nationality, which can impact tones, pitch, timbre of the voice. 
While camera captures facial features, lip and body movement, it has several limitations including privacy issues, limited field of view, and increased energy consumption. 

User position contains important information for interference cancellation. CoS~\cite{jenrungrot2020cone} integrates a binary search of the azimuth direction with separation. A well-known issue of binary search is that multipath signals may be cancelled out in a wider cone that includes the real AoA, which will prevent it from zooming in the right direction and miss the real AoA. 

Ultrasound is also proposed for source separation in \cite{SE-doppler,ultra-noise2}, but specific hardware is not easy to deploy. UltraSE~\cite{UltraSE} emits inaudible sound to the mouth to measure the Doppler shift of lip movement within 0.2 m. 
Our work targets computer users within 1 m away from the speaker/mic.
It is hard to detect lip movements using inaudible acoustic signals in this range. Therefore, we focus on tracking user position using audible and inaudible signals. \cite{WaveEar,Wavoice} use mmWave to sense vocal vibration as a reference of target speech. Our work is complementary and more widely applicable as mmWave deployment is still rather limited whereas speakers and microphones are much more widely available. Moreover, our multi-task learning framework can potentially be applied to other sensing signals.  

There has been significant work in motion tracking in general (\eg, using acoustics~\cite{CAT,MAVL}, WiFi~\cite{rnn_rf_1,decomp_track}, mmWave~\cite{mtrack}, and RFID~\cite{rf-compass, tagoram}). \cite{AAMouse,CAT} use Doppler shift to measure the velocity, \cite{apnea,contactless} use FMCW to measure the distance, and \cite{ubicarse,CSI-loc1} use WiFi Channel State Information (CSI) for localization. There are several algorithms developed for AoA estimation using phase~\cite{rf-idraw}, beamforming~\cite{MVDR}, and subspace methods~\cite{music_fast})). Our work differs from these works by using audible sound to improve multipath profile estimation and using the multipath profile for source separation.

\para{Differences in our work:} 
Our approach makes the following new contributions: (i) We use both user's voice and inaudible tracking signals to generate fine-grained location embeddings, which capture the user's location and rich multipath information. Audible signals and inaudible signals are complementary: audible signals are directly transmitted from the mouth, but have longer wavelength and contain interference and noise; therefore they allow us to estimate the rough AoA region. In comparison, inaudible signals have shorter wavelength and no external interference, but rely on reflected signals, which may come from different body parts. Therefore, inaudible signals are more effective for tracking changes in body position rather than absolute position. Combining the two helps improve the accuracy. 
(ii) We leverage a multi-task learning framework to jointly optimize localization and separation since the two tasks have strong inter-dependency and both depend on the spatial embeddings. 
 (iii) We develop a neural network that uses only a small number of future samples to reduce response time and optimize implementation to provide a real-time guarantee, whereas the existing works mostly focus on the accuracy but few are fast enough for real-time use. 
PHASEN is relatively fast, but its bidirectional structure requires many future samples and increases latency. Adapting PHASEN to use a shorter frame not only requires significant modification but also is insufficient to provide a real-time guarantee since its latency is not reduced proportionally with the frame length due to the lack of batching. In comparison, our work provides real-time source separation. 


%% file: approach.tex
\section{Overview}
\label{sec:approach}

\begin{figure*}
  \centering
  \includegraphics[width=0.8\linewidth]{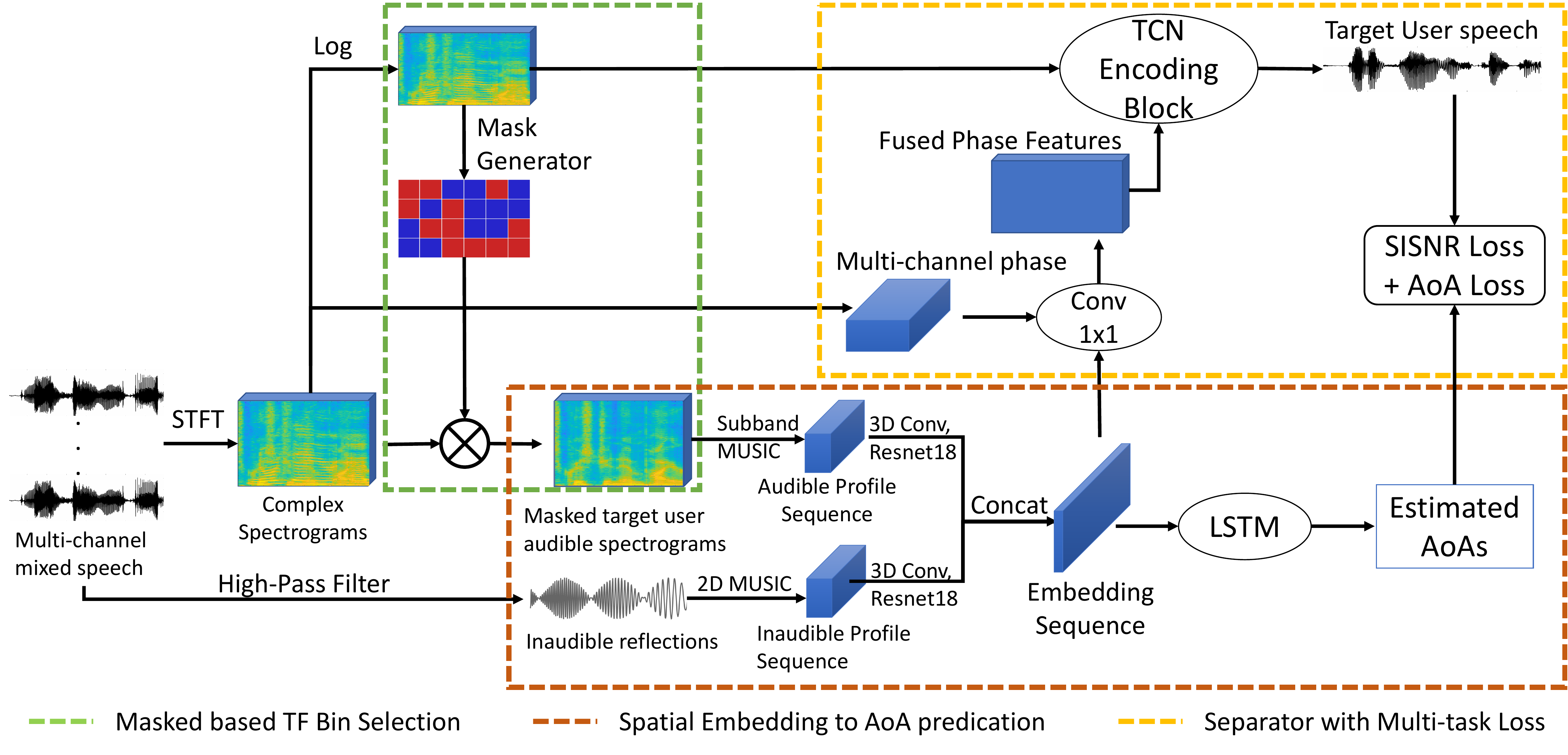}
  \vspace*{-0.2in}
  \caption{\label{fig:overview} It shows the following components: (i) generating masks from the TF bins in audible signals and using the mask to generate MUSIC profiles from speech, (ii) generating spatial embeddings from inaudible tracking sound, and (iii) multi-task learning to jointly separate source and estimate AoA based on the spatial embedding}
  \vspace*{-0.2in}
\label{fig:flow}
\end{figure*}
Consider a user is using a computer with either an internal or external microphone array and speaker to participate in an online meeting. The microphones localize the user's voice. Meanwhile, to improve the tracking accuracy, we also let the speaker emit inaudible tracking sound and use the reflected signals to track the user. Finally, we compute location embeddings from both inaudible and audible signals and feed the embeddings to our source separation neural network to extract the clean speech to send to the Internet. 

Figure~\ref{fig:flow} shows the overall processing. It takes the acoustic signals from a microphone array with a sampling rate of 44.1 kHz, and performs a low pass filter (\ie, 0-8KHz) and a high pass filter (\ie, 18-20 KHz). The outputs from the low-pass and high-pass filters contain audible speech signals and inaudible tracking signals, respectively. We use both signals to generate location embeddings, which will be used together with the audible speech signals under interference and ambient noise to extract clean target signals. 


Our approach consists of four major steps: {\bf (i)} generating location embeddings from audible signals, {\bf (ii}) improving the location embeddings using inaudible signals, {\bf (iii)} using multi-task learning for joint source separation and AoA estimation, and {\bf (v)} speeding up inference. Below we describe each of the steps. 

\comment{
\subsection{Insight from Signal Model Perspective}
We can describe the noisy multichannel speech $\boldsymbol{Y}$ in a traditional signal model. Human speech is broadband signal. Each narrow sub-band of speech can be denoted as 
\begin{equation}
    \boldsymbol{Y}(t,f) = \boldsymbol{H}s(t,f) + \boldsymbol{N}(t, f)
    \label{eq:channelmodel}
\end{equation}
where $s$ represents target clean speech and $\boldsymbol{N}$ is the multichannel noise and interference. $(t, f)$ indicates the time and frequency indices of the acoustic signals in the T-F domain. Various beamformers start from the \ref{eq:channelmodel} to find the optimize weights $\boldsymbol{w}(f)$ to recover the target speech with different optimization objectives, \ie, $\tilde s(t,f) = \boldsymbol{w}^{H}(f)\boldsymbol{Y}(t,f) $. MVDR is one of the most popular beamforming algorithms. Considering no correlation between target speech and interference, the weights of MVDR are given by 
\begin{equation}
    \boldsymbol{w}(f) = \frac{\boldsymbol{\Phi}^{-1}_{Y}(f)\boldsymbol{v}(f)}{\boldsymbol{v}^{H}(f)\boldsymbol{\Phi}^{-1}_{Y}(f)\boldsymbol{v}(f)}
    \label{eq:channelmodel}
\end{equation}

$\boldsymbol{\Phi}^{-1}_{Y}$ stands for the covariance matrix of sub-band noisy speech and $\boldsymbol{v}(f)$ denotes the steering vector of the target speech. $\boldsymbol{v}(f) $ is determined by the AoA and the corresponding frequency. The separation objective is to minimize the difference between beamformed speech $\tilde s$ and target speech $s$. Two branches of research interest are explored in this area. The first branch is to approximate the best $\tilde s$ with given AoAs\cite{}. The other branch is to invest the AoAs with given reference signal $s$\cite{}. Figure \ref{fig:snr_aoa} shows that the more accurate AoAs can result in a better approximation of target speech by MVDR beamforming and vice visa in figure \ref{fig:snr_aoa2}. It searches for optimal AoAs by maximizing the similarity between beamformed signal and given reference speech. The reference speech with higher SNR can result in smaller AoA error. In blind speech separation task, both AoAs and target speech are unknown. While previous work uses a coarse AoA \cite{multi1, multi2}, or an approximate reference speech\cite{UltraSE,audio-visual-cocktail, WaveEar,Wavoice} extracted from other modalities at most, we propose to learn the speech separation and AoA estimation jointly from the received acoustic signals. 
}

\section{Localization using Audible Signals}
\label{ssec:audible}

We first describe how to localize using audible signals. We take the audio signals recorded by a microphone array, and feed them to a low-pass filter (0-8kHz) since human speech is usually below 8 kHz. 
Then we downsample the audio signals to 16 kHz sampling rate, and apply STFT of 512 points to compute the complex spectrogram for different TF bins. We use a hop length of 10 ms in STFT, and a Hamming window of length 32 ms. We then estimate the AoA for each bin and aggregate the estimated AoA across bins using a mask. Below we describe AoA estimation, mask generation, and aggregating AoA profiles using the mask.  



\subsection{AoA Estimation Algorithms} 

A number of algorithms have been developed for AoA estimation, including  phase~\cite{rf-idraw}, beamforming ~\cite{MVDR},
 and MUSIC~\cite{music}. We use MUSIC for its high accuracy. In MUSIC, we compute the auto-correlation matrix $R = x^{H} x$, where $x$ is the received signal and $x^H$ is conjugate transpose of $x$. We then perform eigenvalue decomposition on $R$, and sort the eigenvectors in a decreasing order of the corresponding eigenvalues. The signal space consists of the first $M$ eigenvectors. The noise space, denoted as $R_N$, consists of the remaining eigenvectors. Let $\theta$ and $\phi$ denote the azimuth and elevation angles, respectively. We derive the pseudo-spectrum of the mixed signals based on $R$ as
$p(\theta,\phi)=\frac{1}{a(\theta,\phi)^{H} R_N R_N^{H} a(\theta,\phi)}$, where
$R_N^{H} \cdot a(\theta_0,\phi_0) = 0$.
In the free space, AoA can be estimated by searching for a peak in the pseudo-spectrum. Multipath makes AoA estimation more challenging by introducing multiple peaks and merged peaks. Ambient noise and interference can further complicate the issues by adding false peaks. Therefore, it is hard to get a reliable AoA estimate in general.




\subsection{Mask Based TF bin selection} 

We improve the MUSIC accuracy by carefully selecting the frequency bins in the audible signals for aggregation to minimize the impact of noise and interference.  Specifically, since MUSIC assumes narrowband signals, we apply STFT to the audible signals to generate TF bins, where each bin occupies 31.25 Hz and  10 ms. We then perform MUSIC on each TF bin, and aggregate the MUSIC profiles across TF bins.

A natural approach to aggregate the MUSIC spectrum across different frequency bins is to sum up the MUSIC profiles from all TF bins and select the angle corresponding to the highest peak. However, not all TF bins contain the target user's speech due to the sparsity of human speech over the frequency band~\cite{Wang2016}. 
Therefore, it is important to select the TF bins that contain strong Signal to Noise Ratios (SNR) from the target user. It is challenging to select the TF bins by just analyzing the power and phase because, unlike tracking signals, human speech is out of our control and hard to predict. Moreover, some TF bins may contain significant ambient noise and interference, so we cannot simply select the TF bins solely based on the overall magnitude, but should select the TF bins with high SNR from the target user. 





\para{Mask generation:} A number of approaches have been proposed to generate TF masks for speech enhancement. A few works estimate the amplitudes of audio spectrogram (\eg, real valued ideal binary mask (IBM)~\cite{IBM,IBM2} and ideal ratio mask (IRM)~\cite{IRM}). 
\cite{Wang2016,PHASEN} develop DNN based approaches to generate amplitude and phase masks. Most of these methods focus on combating noise. 


We use IBM to select appropriate TF bins for the target user from mixed noisy complex spectrogram. IBM is a method for speech separation based on deep neural network~\cite{IBM2}. 
Even though IBM is not the best method for source separation, it is a good fit for selecting the TF bins dominated by the target speaker. Other mask-based methods apply linear translation to the original TF bins, which introduces phase distortion and degrades mask generation. 


IBM is based on the sparsity of human speech (\ie, the number of non-silent TF bins from a speaker tends to be small). It determines a binary mask for each TF bin, where 1 means the target signal dominates interference and 0 otherwise. It takes a downsampled audio signal and decomposes it into 2D TF bins. Then it extracts several features, such as autocorrelation of a filter response, autocorrelation of an envelop of filter response, and cross-channel correlation. Next, it performs clustering based on these extracted features (\eg, cluster into a target stream and an interference stream), and tag each TF bin with either target dominant ("1") or interference dominant ("0") based on similarity with the clean target signal (spoken at a different time), which is also an input. We use the clean target signal, which is location independent and can be collected only once during user account creation. Since an online meeting requires a user to sign in, it is reasonable to assume the target user is known. To extend to the scenario where the target user is not known, one can apply speaker recognition algorithms (\eg, \cite{speaker-recog}) and use the clean speech from the identified user for mask generation.

We train the IBM mask estimator using Deep Cluster ~\cite{deepcluster}, which is a general and robust method to estimate the mask. It takes log power spectrogram (LPS) (\ie, the log power of received signal across 256 positive frequency bins) as the input to estimate an initial binary mask of the target user. We get the pre-trained model using the LibriMix~\cite{librimix} data along with our own testbed traces described in Section~\ref{ssec:eval-method}. The binary mask is a coarse estimate of effective TF bins, but it maintains phase information and is useful for selecting the TF bins for further analysis.  We estimate binary masks for all microphones. To minimize interference, we select the TF bins for AoA estimation only when the masks from all microphone channels are 1s. In this way, we effectively remove the TF bins with large interference and noise.




\para{Applying mask to MUSIC:} We apply the MUSIC algorithm to the TF bins with masks. Then we concatenate the MUSIC spectrum from all frequencies together. The output profile is represented as a 2D matrix of size $M_{f} N_{a}$ across different frequencies, where $M_{f}$ denotes the number of frequency bins and $N_{a}$ denotes the number of angles. $M_{f}$ is set to 103 frequency bins (equally spaced from 800Hz to 4KHz for human speech), and $N_{a}$ is set to 181 (spanning 0 degree to 180 degree with 1 degree apart) in our evaluation. This will be further combined with the output from inaudible signals for generating location embeddings. 



\section{Leveraging Inaudible Sensing}
\label{ssec:inaudible}

\subsection{Motivation for using inaudible signals} 

\begin{figure}[t]
\centering
\begin{minipage}{0.25\textwidth}%
\centering
\includegraphics[width=\columnwidth]{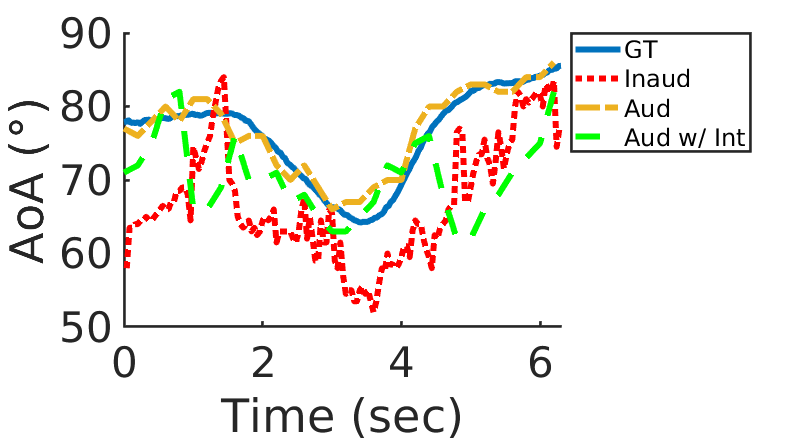}
\caption{Plausible AoAs estimated from audible band and inaudible band}
\label{fig:aoa_err_compare}
\end{minipage}
\begin{minipage}{0.22\textwidth}%
\centering
\includegraphics[width=\columnwidth]{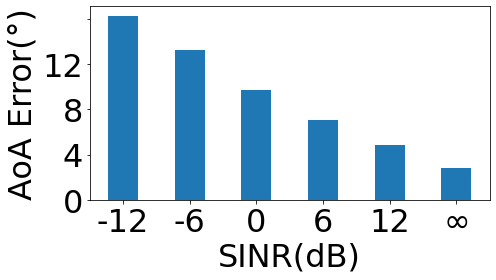}
\caption{AoA error under different SINR}
\label{fig:aoa_err_SNR}
\end{minipage}
\vspace*{-0.15in}
\end{figure}


Estimating multipath profile solely based on speech has several limitations: (i) Audible signals may contain significant interference and ambient noise, which results in significant AoA errors. Figure~\ref{fig:aoa_err_compare} depicts the AoA estimation drifts from the ground truth with the interference in audible band. To quantify the impact of SNR, we apply MUSIC to the audible signals under different SNR. We scale the magnitude of interference and mix up with target speech. As shown in Figure~\ref{fig:aoa_err_SNR}, adding interference increases the AoA error significantly. For example, SNR = -6 dB increases the AoA error over no interference by $10.45^{\circ}$, and SNR = - 12 dB increases the error by $13.53^{\circ}$. While using masks removes a significant amount of interference, the removal is not perfect. (ii) Most energy in audible signals concentrates in low frequency (\eg, below 2 kHz), which has a large wavelength and leads to low resolution in AoA for a fixed microphone separation. (iii) Typically a relatively large time window (\eg, hundreds of ms) is used to analyze audible signals in order to ensure there is enough energy from the target speaker for many time windows. This limits the update rate of multipath profile estimation. 

In comparison, inaudible signals have shorter wavelength, hence higher precision. Moreover, they also allow us to track at a much higher frequency (\eg, tens of ms), which enables us to adapt more quickly to the changing user position. However, in the typical usage scenarios where the speaker and mics are on the desk, inaudible signals are mostly reflected by the user's body instead of mouth. Therefore, inaudible signals mainly track the body movement instead of lip movement. Besides, there are multiple reflections from the human body. Fortunately, since body movement will induce mouth movement,  inaudible signals can capture the motion dynamics to help correct inaccurate AoA estimation from noisy audible band. Figure ~\ref{fig:aoa_err_compare} shows an example of inaudible AoA trace: it differs from the ground truth AoA of the mouth, but has a similar trend to the mouth movement. There are multiple AoA candidates from inaudible reflection and they have similar movement trends, which are useful for tracking and source separation. We develop a neural network to exploit the features extracted from both audible and inaudible signals. 



\subsection{Feature extraction with 2D MUSIC} 
We let a speaker (either internal or external) on a computer transmit periodic FMCW chirps, whose frequency increases linearly from $f_{min}$ to $f_{max}$ during each period $T$. This yields a transmission signal $u_t(t) = cos(2\pi f_{min} t' + \frac{\pi B t'^2}{T})$, where $t' = t-nT$. 

After going through the channel with the propagation delay $t_d$ and attenuation $\alpha$, the received signal becomes $u_r(t) = \alpha cos(2\pi f_{min} (t'-t_d) + \frac{\pi B (t'-t_d)^2}{T})$. 
We collect the received signal reflected by the user from a microphone array, and apply the MUSIC algorithm to derive the pseudo spectrum. In order to get a more fine-grained pseudo spectrum, we use 2D MUSIC~\cite{2dmusic} to generate a distance and AoA profile $P(d,\theta)$, where $d$ and $\theta$ denote the distance and Azimuth angle, respectively. 



In the free space, the peak in the 2D MUSIC profile indicates the target AoA and distance. In practice, signals traverse over multipath, which may introduce multiple peaks. If the two paths are too close, their resulting peaks can get merged and the highest point in the peak may not correspond to the target location. Noise and Doppler shift resulting from movement complicate the 2D MUSIC profiles by adding false peaks and distorting the real peak. 
Therefore, instead of selecting a single peak for AoA and distance estimation, which may introduce error that is hard to recover later, we feed the complete 2D MUSIC profile from inaudible tracking along with the MUSIC profile from voice signals to generate location embeddings. 

\subsection{Generating Spatial Embeddings}

\label{ssec:combine}
We regard the MUSIC profile sequence as a sort of video sequences. Thus, we take advantage of state-of-the-art modules in video processing.
We apply 3D convolution~\cite{3dconv} with kernel 5 × 7 × 7 and ResNet-18~\cite{resnet} to MUSIC profiles generated from audible and inaudible signals to effectively learn spatio-temporal features. 
The small temporal window of 3D kernels helps filter out noisy patterns based on neighbor frames. It is followed by batch normalization and ReLu activation. Then output features are fed to ResNet18 to encode profile embeddings with 512 dimensions. The concatenated embeddings from audible and inaudible profiles will be directed to the source separation.

A sequence of embeddings are extracted from the MUSIC profile sequence. We feed them into both LSTM and source separation network. LSTM takes these embeddings to estimate AoAs because they include important spatial information from audible and inaudible signals over the recent time window. The 3D convolutional layers use temporal information in small time windows at early stage, while the LSTM can leverage a much longer sequence. Moreover, audible embedding and inaudible embedding complement each other due to the correlation between the mouth movement and body movement. The memory unit in LSTM helps track a long-term movement. Our objective is to minimize the L1 loss between the estimated AoA ($\hat{AoA}$) and ground truth AoA ($AoA$), denoted as  $L_{AoA} = ||AoA - \hat{AoA} ||_1$. We use LSTM to estimate the AoA based on the audible and inaudible location embeddings. The LSTM has one LSTM layer with 128 nodes, which is followed by two fully connected layers. The first layer has 64 ReLU nodes, and the second layer outputs a single AoA. 

\section{Multi-task Learning for Source Separation}
\label{ssec:separate}

So far, we have focused on localizing the target user. Next, we consider how to leverage the location information for source separation. Due to the strong inter-dependency between tracking and source separation, we apply a multi-task learning framework to jointly estimate the location and separate the source. 


\subsection{Learnable Pre-mask} 
Previous works~\cite{multi1} develop a pre-mask to take into account of AoA $\theta$. It first forms a steering vector $\boldsymbol{e}_{\theta}(f)$ based on the AoA,  and computes the cosine distance between the steering vector and the complex values in each TF bin as follow:  
\begin{equation}
    A(t,f) = \sum_{k=2}^M \frac{\boldsymbol{e}_{\theta, k}\frac{\boldsymbol{Y}_k(t,f)}{\boldsymbol{Y}_1(t,f)}}{|\boldsymbol{e}_{\theta, k}\frac{\boldsymbol{Y}_k(t,f)}{\boldsymbol{Y}_1(t,f)}|}
    \label{eq:premask}
\end{equation}
where $\boldsymbol{Y}$ is the complex spectrogram, $M$ is the number of microphone channels, and $k$ is the microphone index starting from the second microphone as the steering vector is normalized to the first microphone. $A(t,f)$ represents the pre-mask value to a TF bin.
The pre-mask indicates the probability of a TF bin dominated by the source coming from the given AoA. Intuitively, the pre-mask lets the network beamform towards a given direction. Pre-mask improves over traditional linear beamformers (\eg, MVDR~\cite{MVDR} and LCMV~\cite{LCMV}) by using a DNN-based non-linear filter, so it has better spatial discrimination and interference cancellation. 

The pre-mask assumes the input AoA is accurate. In our context, the AoA estimation can be erroneous due to interference, ambient noise, and multipath propagation. Moreover, not only the direct path but also reflected paths are important for source separation because the overall received phase is the result of all multipath. A single AoA estimate does not provide complete spatial information of the target speaker. Therefore, we propose to fuse our spatial embeddings with the mixed phase from the complex spectrogram to learn a better spatial pre-mask. 


For each microphone and TF bin, there is a 512-long embedding from audible profiles and another 512-long embedding from inaudible profiles. 
These embeddings are concatenated and processed by 1x1 convolutional layer followed by a layer normalization and PReLU. The output of each TF contains a spatial feature map. It is concatenated with LPS and fed into Temporal Convolution Network (TCN)~\cite{Conv-TasNet}. TCN outputs a mask, which can be applied to the mixture complex spectrogram to generate the target complex spectrogram. Then we perform inverse short-term Fourier Transform to estimate the target signals.

\begin{figure}[t]
\centering
\begin{minipage}{0.23\textwidth}%
\centering
\includegraphics[width=0.98\columnwidth]{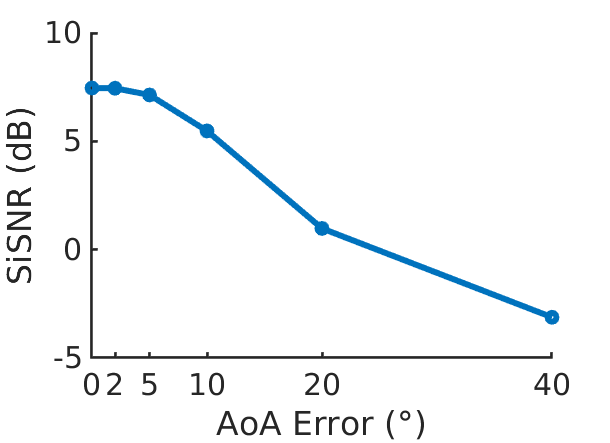}
\caption{The reduced AoA error improves the performance of MVDR beamformer.}
\label{fig:snr_aoa}
\end{minipage}
\begin{minipage}{0.23\textwidth}%
\centering
\includegraphics[width=0.98\columnwidth]{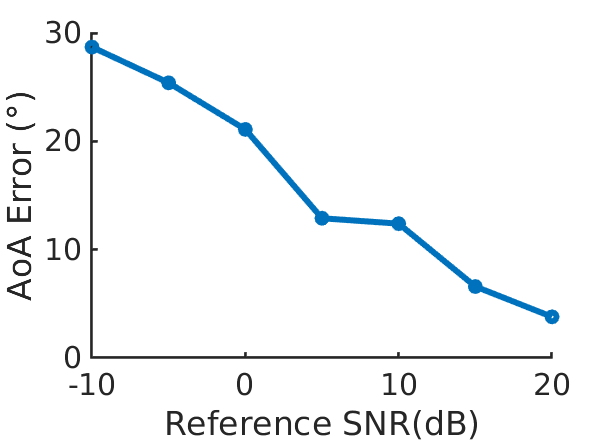}
\caption{The increased SNR of reference speech decreases the AoA error.}
\label{fig:snr_aoa2}
\end{minipage}
\vspace*{-0.2in}
\end{figure}

\subsection{Multi-task learning for Separation and AoA Estimation} 

We develop a multi-task learning for source separation and AoA estimation. Our approach is motivated by the strong inter-dependency between the two tasks. We consider MVDR, one of the most popular beamforming algorithms, as an example. Assuming no correlation between the target speech and interference, the weights of MVDR are given as follow:
\begin{equation}
    \boldsymbol{w}(f) = \frac{\boldsymbol{\Phi}^{-1}_{Y}(f)\boldsymbol{v}(f)}{\boldsymbol{v}^{H}(f)\boldsymbol{\Phi}^{-1}_{Y}(f)\boldsymbol{v}(f)}
    \label{eq:channelmodel}
\end{equation}
where $\boldsymbol{\Phi}^{-1}_{Y}$ is the covariance matrix of sub-band noisy speech and $\boldsymbol{v}(f)$ is the steering vector of the target speech. The goal of source separation is to minimize the difference between beamformed speech $\tilde s$ and target speech $s$. Figure \ref{fig:snr_aoa} shows that more accurate AoA estimation yields better source separation and an improved source separation reduces AoA estimation error. In blind speech separation task, both AoAs and target speech are unknown. Therefore, we propose to learn the speech separation and AoA estimation jointly from the received acoustic signals. 
\comment{
We can describe the noisy multichannel speech $\boldsymbol{Y}$ in a traditional signal model. Human speech is broadband signal. Each narrow sub-band of speech can be denoted as 
\begin{equation}
    \boldsymbol{Y}(t,f) = \boldsymbol{H}s(t,f) + \boldsymbol{N}(t, f)
    \label{eq:channelmodel}
\end{equation}
where $s$ represents target clean speech and $\boldsymbol{N}$ is the multichannel noise and interference. $(t, f)$ indicates the time and frequency indices of the acoustic signals in the T-F domain. Various beamformers start from the \ref{eq:channelmodel} to find the optimize weights $\boldsymbol{w}(f)$ to recover the target speech with different optimization objectives, \ie, $\tilde s(t,f) = \boldsymbol{w}^{H}(f)\boldsymbol{Y}(t,f) $. MVDR is one of the most popular beamforming algorithms. Considering no correlation between target speech and interference, the weights of MVDR are given by 
\begin{equation}
    \boldsymbol{w}(f) = \frac{\boldsymbol{\Phi}^{-1}_{Y}(f)\boldsymbol{v}(f)}{\boldsymbol{v}^{H}(f)\boldsymbol{\Phi}^{-1}_{Y}(f)\boldsymbol{v}(f)}
    \label{eq:channelmodel}
\end{equation}
where $\boldsymbol{\Phi}^{-1}_{Y}$ is the covariance matrix of sub-band noisy speech and $\boldsymbol{v}(f)$ is the steering vector of the target speech. $\boldsymbol{v}(f) $ is determined by the AoA and the corresponding frequency. The separation objective is to minimize the difference between beamformed speech $\tilde s$ and target speech $s$. Two branches of research interest are explored in this area. The first branch is to approximate the best $\tilde s$ with given AoAs\cite{}. The other branch is to invest the AoAs with given reference signal $s$\cite{}. Figure \ref{fig:snr_aoa} shows that the more accurate AoAs can result in a better approximation of target speech by MVDR beamforming and vice visa in figure \ref{fig:snr_aoa2}. It searches for optimal AoAs by maximizing the similarity between beamformed signal and given reference speech. The reference speech with higher SNR can result in smaller AoA error. In blind speech separation task, both AoAs and target speech are unknown. While previous work uses a coarse AoA \cite{multi1, multi2}, or an approximate reference speech\cite{UltraSE,audio-visual-cocktail, WaveEar,Wavoice} extracted from other modalities at most, we propose to learn the speech separation and AoA estimation jointly from the received acoustic signals. 
}

A common learning objective in the existing separation network is to maximize SiSNR~\cite{Conv-TasNet}. Let $\hat{x}$ denote the estimated signal and $x$ denote the clean reference signal. We compute SiSNR as follow: 
\begin{equation}
    SiSNR = 10 log_{10} \frac{||x_{target}||^2}{||e_{noise}||^2}
\end{equation}
where $x_{target} = \frac{<\hat{x},x>x}{||x||^2}$,  $e_{noise} = \hat{x}-x_{target}$. By normalizing $x$ and $\hat{x}$ to zero mean, we ensure scale invariant. The loss function is defined as $L_{SiSNR} = -SiSNR$. Following the existing work (\eg, \cite{UltraSE}), the target and interference signals are measured separately and added up to simulate interference. Therefore, SiSNR can be computed based on their values.

Unlike the existing works that optimize only SiSNR, we develop a novel multi-task learning framework to jointly learn speech separation and AoA. Multi-task learning trains ML models for multiple tasks simultaneously using a shared structure. The idea of multi-task learning is that internal representations learned for one task can be helpful for the other tasks, and vice versa. 

Our key observation is that the spatial embedding can benefit both speech separation and AoA estimation. An accurate embedding enables LSTM to accurately estimate the AoA. It also provides good hints for TF bins, which will be fused with spatial knowledge and target speaker direction.


Another important insight is that jointly learning the separation and AoA can reinforce the network to learn the AoA instead of treating the AoA as the fixed input, which prevents the gradient from propagating back and contributing to the training task. In comparison, when the AoA is set to be learnable together with separation, the mixed phase can contribute to the learning objective. By fusing the phase and embedding, the learned phase is more consistent with the separated source. The phase is represented as a 2D tensor, which represents the phase over different microphones and frequencies. We use more convolutional layers to learn the phase using the following objective for training:
$$ L = L_{SiSNR} + \lambda L_{AoA}$$
where $\lambda$ is a relative weighting factor and set to 0.5 in our evaluation. 

\begin{figure}[t]
  \centering
  \vspace*{-0.2in}
  \includegraphics[width=0.85\columnwidth]{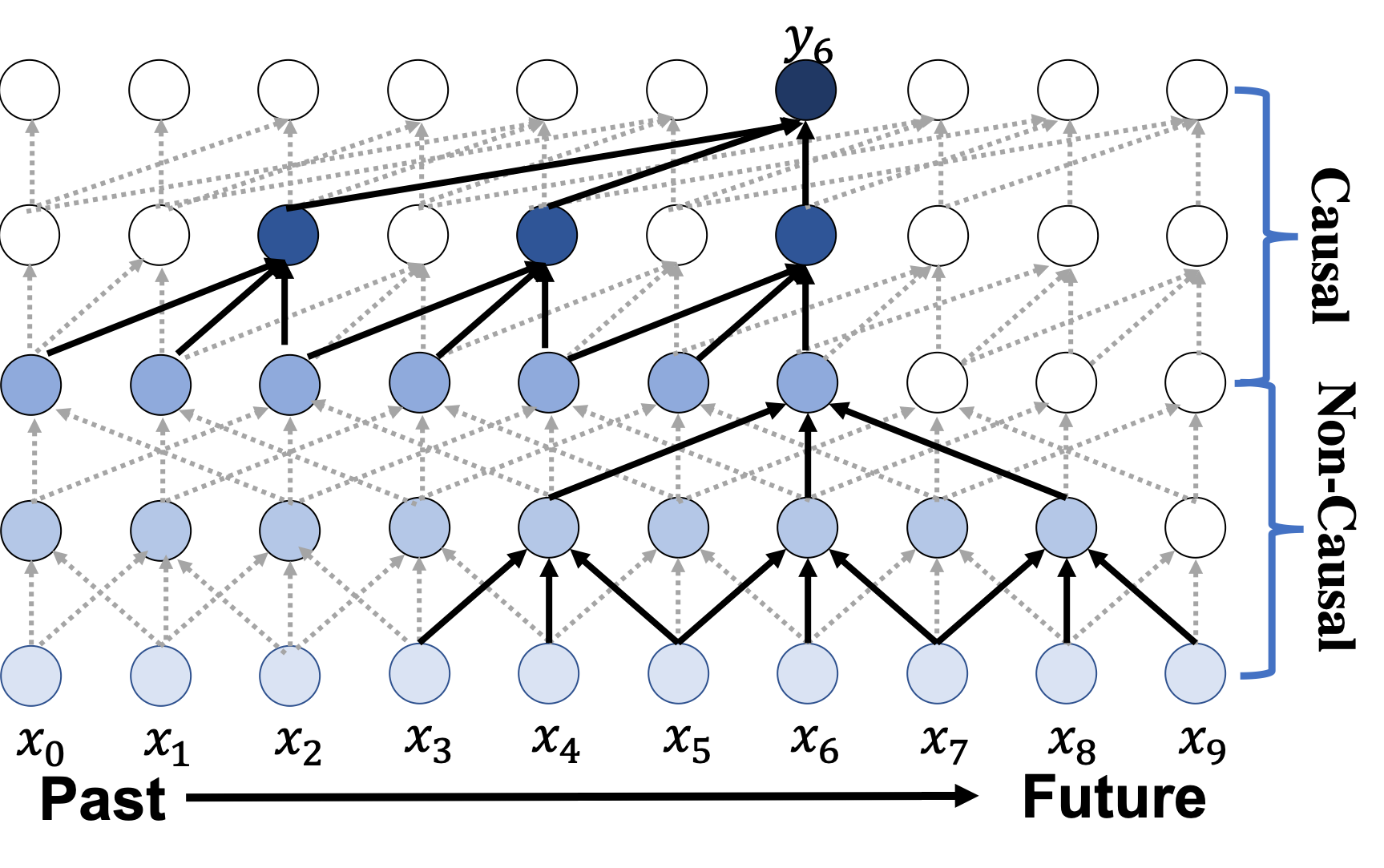}
  \vspace*{-0.2in}
  \caption{\label{fig:conv}Stack causal and non causal convolution layers.}
  \vspace*{-0.2in}
\end{figure}

\section{Real-Time Model Design and Implementation}

\subsection{Real-Time Model Design} While several source separation schemes claim to achieve real-time inference because their processing time is shorter than the audio duration (\eg, 1.71s processing time for 4s audio in Conv-TasNet\cite{Conv-TasNet}), this is insufficient to ensure real-time. Another requirement is that the output should be generated without much delay. This has significant implication on the neural network structure. In particular, existing works use non-casual convolutional layers, which do not distinguish between past and future input samples and require many future samples (\eg, 1.28 seconds in Conv-TasNet). 

In order to support real-time processing, we only use a small look-ahead window (\eg, 90 ms in our implementation). We show that even a small look-ahead is sufficient to yield good performance. Moreover, we pay special attention to causal vs. non-casual convolution. The top 2 layers are causal convolution in Figure ~\ref{fig:conv}. They only require previous samples. They perform worse than the non-causal layers because they do not use the future samples. The bottom two layers are non-causal convolution which is common in most network architecture. They need future samples to perform convolution operation. The stack of non-causal layers increases the perception field of future samples exponentially. In order to achieve high accuracy while limiting the latency, we configure the first 2 convolutional layers of each TCN block as non-causal with a small look-ahead and the other layers as causal. Figure~\ref{fig:conv} shows an example: at timestamp 6, the first two non-casual layers use samples up to timestamp 9 and the next two casual layers only use samples up to the current timestamp 6.




\subsection{Real-Time Model Implementation} 

We implement the \sysname model in Pytorch~\cite{pytorch}. We use the Adam optimizer with an initial learning rate of $0.0001$. We apply a multi-step scheduler to drop the learning rate by 50\% at epochs 40 and 75. The maximum training epoch is 150 but it will stop early if there is no more improvement for 10 epochs. \sysname has 20.2M parameters in total. 

To provide real-time guarantee, we cannot wait to accumulate a few second audio before processing, but process more frequently (\eg, at least once every 150 ms). But the processing time does not reduce proportionally with the reduced input size due to the lack of batching opportunities. Hence how to achieve real-time ML based source separation remained open. 

To speed up processing, our system processes audio every 90 ms. We introduce a cache tensor for each block to cache the previously computed intermediate result in the neural network and reuse it in the next round. Moreover, we use Microsoft Onnxruntime~\cite{onnx} to significantly speed up the inference. Since certain operations are not supported in Onnxruntime, we replace these operations with similar but supported operations. Together, the resulting system achieves real-time processing -- it processes 90 ms audio within 42 ms, which yields the total latency of $90+42 = 132$ ms (within the 150 ms real-time requirement).

%% file: eval.tex
\section{Evaluation}
\label{sec:eval}

In this section, we first present our evaluation methodology and then describe our performance results.

\subsection{Evaluation Methodology}
\label{ssec:eval-method}
 
 \begin{figure}[t]
\centering
\begin{minipage}{0.18\textwidth}%
\centering
\includegraphics[width=0.98\columnwidth]{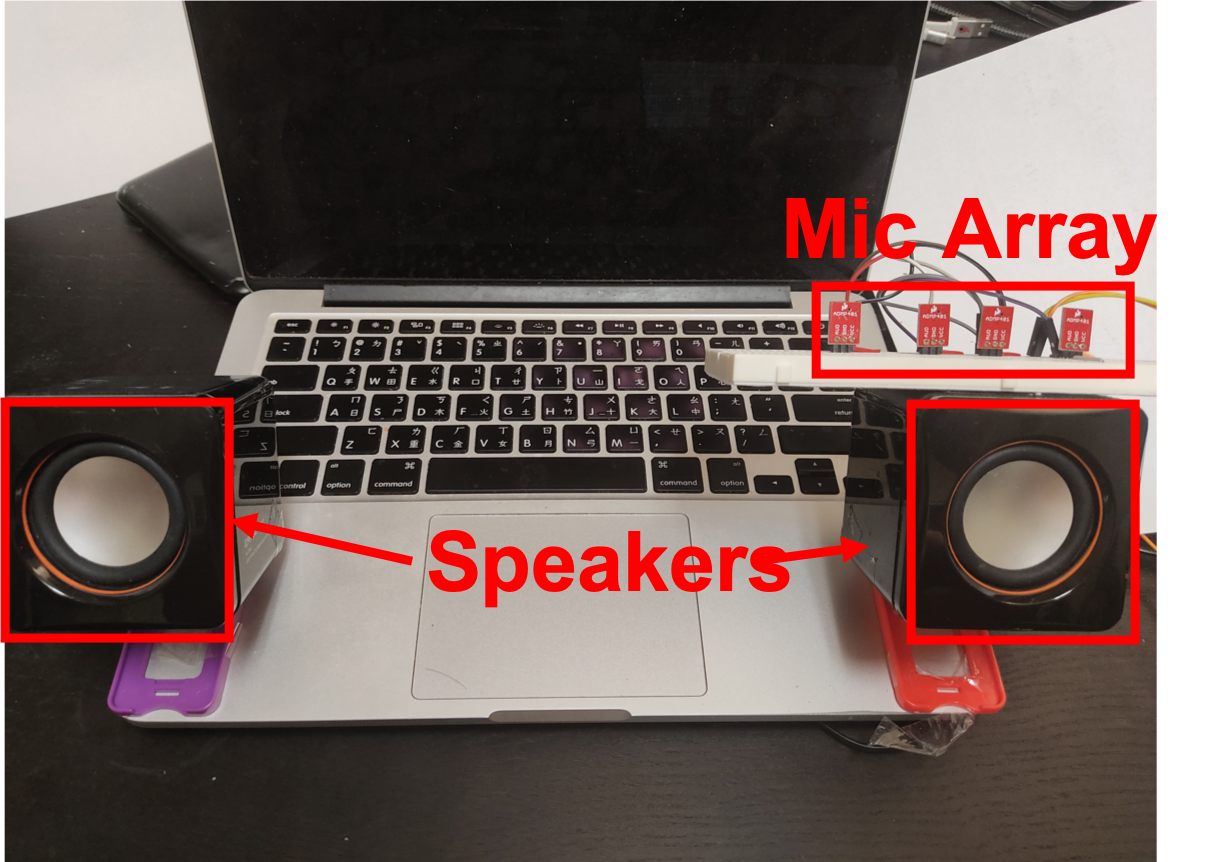}
\caption{Platform setup}
\label{fig:setup}
\end{minipage}
\begin{minipage}{0.28\textwidth}%
\centering
\includegraphics[width=0.98\columnwidth]{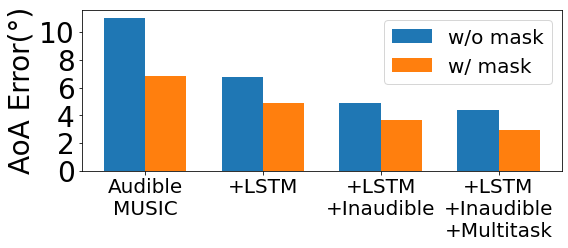}
\caption{AoA Error of Different Variants .}
\label{fig:aoa_err}
\end{minipage}
\vspace*{-0.2in}
\end{figure}



\para{Setup:} As shown in Figure~\ref{fig:setup}, we connect a laptop with a Bela platform~\cite{Bela} attached with a pair of speakers and four microphones. The microphones form a linear array spanning 8 cm with non-uniform space between them. Their positions are [0, 3 cm, 5 cm, 8 cm]. A similar setup is used in ~\cite{Rtrack}. The processing is done on the laptop using Pytorch. We train the model on the GPU of NVIDIA GTX 2080ti. We run inference on both the desktop and Macbook.

\para{Data collection:} There are no open source multi-channel recordings for our evaluation. Therefore, we collect the data on our own. We will release the data to the public. Our data include multi-channel inaudible and audible audio signals. We let dual speakers both transmit periodic FMCW chirps from 18-20 kHz with a period of 40 ms and a sampling rate of 44.1 kHz. To avoid interference, the two speakers transmit the same chirp with 20 ms difference in the starting time. The volume of the speakers is set to a little less than the maximum to avoid signal distortion. We set another headset microphone to record the reference clean speech.  

We collect 20 users' speech using our setup. Among them, there are 8 females and 12 males. There are two kids: 8 and 13 year old, and the rest are adults between 22 -- 59 year old. Each user speaks for 10 minutes - 1 hour in total. We let the users present slides or read books or papers to mimic online conferences. This is an easy way for users to generate continuous speech. The target user is 0.2--0.7 m away from the microphone. The users move naturally during the trace collection. For example, they sometimes lean towards or away from the computer, move side to side, or turn their heads. We collect the data from different environments (\eg, lab, living room, study room, cubicle, and conference room). The environments have different multipath, which affects both AoA estimation and source separation. 

We also separately record interference by letting an external speaker play a random set of speech from Librispeech~\cite{librispeech}, which contains more than 1K speakers and 26K English sentences lasting 1000 hours. We place the interfering speaker inside or outside the room where the target user is located. When the interfering speaker is inside the room, (s)he is a couple of meters away. We augment the real traces by scaling the SNR of the target signals from -6 dB to 6 dB. Moreover, we use gpuRIR~\cite{gpurir} to simulate realistic interference and noise in multi-channel scenarios by estimating and applying Room Impulse Responses (RIR) to clean speech from Librispeech and noise in WHAM!~\cite{wham} respectively. 
WHAM! collects ambient noise in non-stationary environments, such as coffee shops, restaurants, and bars. These sounds are generated by humans, musical instruments, and vehicles. Adding such noise to the background interference makes it even more difficult to extract clean signals. 

\para{Dataset Preparation:} We mix the audio segments containing the target speaker's speech and inaudible FMCW reflection with different types of interference and background noise. We add different interference and noise to each target user's speech. We vary the amount of interference and background noise according to the required SNR. The number of interfering users is uniformly distributed between 0 and 3, and the SNR is uniformly distributed between $[-6, 6]$ dB. In total, the training data is generated from 16 users' speech. It contains 30K segments of mixed audio signals, where each segment lasts for 4 seconds and the total training data lasts for 31 hours. The testing data contains 6K segments generated from 4 user. Following the common practice, we vary the user in the testing dataset and use the remaining user for training.  Both training and testing have real recording samples from all environments. Interference and noise are from training split and test split of LibriSpeech and WHAM! respectively. In addition, we also evaluate how our model generalizes to a new environment that is not present in the training traces. 


\para{Performance metrics:} Following the existing works (\eg, \cite{UltraSE}), we use several metrics to quantify the performance of source separation:
(i) {\bf SiSNR} prevents unfair impact of the rescaled signals~\cite{TasNet1}; (ii) {\bf Short-time objective intelligibility measure (STOI)} quantifies intelligibility of speech~\cite{STOI}; (iii) {\bf Perceptual Evaluation of Speech Quality (PESQ)~\cite{PESQ}} is designed to quantify the quality of processed speech, and its score ranges from 1 to 5. Higher values in the above metrics indicate better speech quality.

In addition, we also report AoA estimation errors. We measure the ground truth AoA using the Intel RealSense L515~\cite{keselman2017intel}. To ensure the RealSense gets accurate AoA, we place it in line with the microphone array and provide good lighting conditions. 

\para{Baselines:} We compare \sysname with the following state-of-the-art approaches: (i) Conv-TasNet~\cite{Conv-TasNet}: It is one of the best speech separation approaches using single-channel speech. It is also one of the most widely used baselines due to its open-source. (ii) PHASEN~\cite{PHASEN}: It is a denoising network using two streams to improve phase estimation. 
UltraSE~\cite{UltraSE} shows that \cite{Conv-TasNet} and \cite{PHASEN} are the best baselines that only use speech for source separation. All schemes are trained using the same data as our approach. 

We did not compare with UltraSE~\cite{UltraSE}, which targets phone users. 
We target computer users for online meetings, which is complementary to UltraSE and also more common than smartphone users  (\eg, \cite{video-conf-statistics} reports the majority of users use computers for online meetings). Moreover, UltraSE requires the phones' speaker/mic to face the user's mouth and within 20 cm, which is even less common as most speakers/mics on the phone face bottom instead of users.  Meanwhile, our measurement shows that  the headset Sennheiser DK-2750 improves SiSNR by 8.91 dB over the internal microphones of the laptop under interference. In comparison, our software only solution provides higher SiSNR, and hence is more attractive.  

\subsection{Performance Results}
\label{ssec:eval-perf}

\subsubsection{Micro Benchmark}

We first present micro benchmarks, where we compare different variants of our own algorithm. In the micro-benchmark, we use one interfering user and background noise setup. Unless otherwise specified, we use the data collected from all environments, 20 users, and SNR range of -6dB to 6dB as the default settings. 

\para{Impact of AoA estimation algorithms:} We first compare different variants of our AoA estimation.  Figure~\ref{fig:aoa_err} plots the average AoA estimation error. The basic method is to apply standard wideband MUSIC to estimate the AoA (\ie, applying MUSIC to each frequency band and summing up the results across all bands). We then augment the method with various enhancements.  As it shows, each of our enhancements, namely mask, LSTM, inaudible tracking, and multi-task learning, helps improve the AoA error. Using mask reduces the AoA error by $0.5^{\circ} - 4.2^{\circ} $ across different cases  by removing the most noisy TF bins to prevent generating incorrect MUSIC spectrum. Using LSTM brings an additional $1.9^{\circ}$ improvement over using the MUSIC profile in a single period since it leverages the inherent temporal locality in the movement. Using inaudible tracking further reduces the AoA error by $1.2^{\circ}$ through overcoming audible noise and interference and updating the location more frequently. Finally, multi-task learning improves the AoA by another $0.7^{\circ}$ through jointly optimizing the source separation and AoA estimation. Putting everything together, we achieve $3.8^{\circ}$ AoA estimation error. 

 

\begin{figure*}[t]
\centering
\vspace*{-0.15in}
\begin{minipage}{0.24\textwidth}%
\centering
\includegraphics[width=0.95\columnwidth]{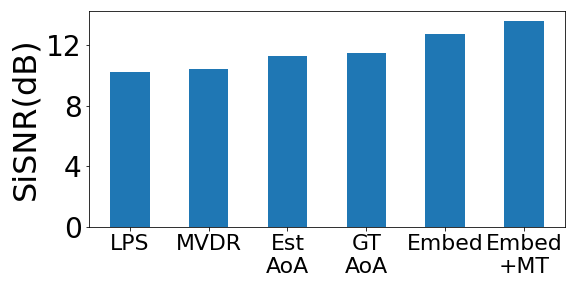}
\vspace*{-0.15in}
\caption{Different model structure}
\label{fig:SiSNR_res}
\end{minipage}
\begin{minipage}{0.24\textwidth}%
\centering
\includegraphics[width=0.95\columnwidth]{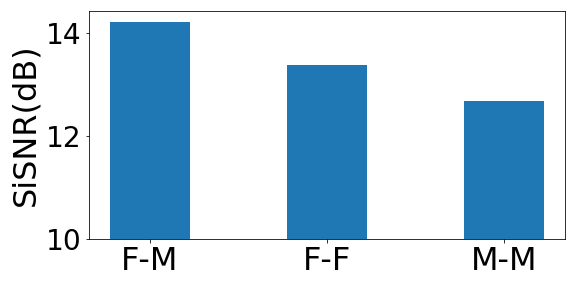}
\vspace*{-0.15in}
\caption{Different gender combinations}
\label{fig:user_SiSNR}
\end{minipage}
\begin{minipage}{0.24\textwidth}%
\centering
\includegraphics[width=0.95\columnwidth]{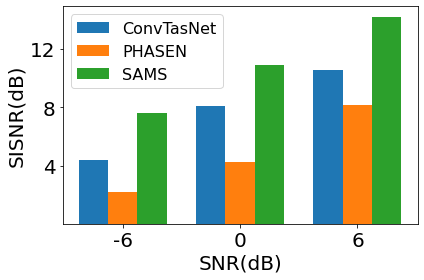}
\vspace*{-0.15in}
\caption{Different SNR}
\label{fig:snr_SiSNR}
\end{minipage}
\begin{minipage}{0.24\textwidth}%
\centering
\includegraphics[width=0.95\columnwidth]{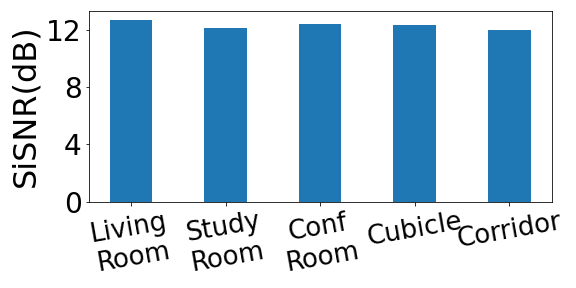}
\vspace*{-0.15in}
\caption{Different Environments}
\label{fig:room}
\end{minipage}
\vspace*{-0.15in}
\end{figure*}

\para{Impact of different separation algorithm:}  Next we compare with the following source separation algorithms using the same set of data: 
    (i) LPS~\cite{lps}: It uses only a single channel LPS of raw mixed audio signals for source separation.
    (ii) MVDR~\cite{MVDR}: We estimate the AoA by applying MUSIC to audible signals, and use MVDR to beamform towards the AoA. 
    (iii) Est AoA: We estimate the AoA using both audible and inaudible AoA and use our source separation DNN to extract the target signals. It differs from our final algorithm in two aspects: (i) it uses AoA for source separation instead of location embeddings, and (ii) it disables multi-task learning.
    (iv) GT AoA: It applies our source separation network to the ground truth AoA. Like the above estimated AoA, it disables multi-task learning and uses AoA instead of location embeddings in source separation. 
    (v) Embed: It uses spatial embeddings but disables multi-task learning in our final approach. 
    (vi) MT: This is the final approach that uses both location embeddings and multi-task learning. 



As shown in Figure~\ref{fig:SiSNR_res}, on average, LPS yields 10.26 dB SiSNR by leveraging the magnitude information to learn the acoustic model of speech. MVDR beamforming achieves 10.43 dB SiSNR by leveraging the AoA estimate from audible signals. The improvement is limited due to the limited accuracy of AoA estimation. By leveraging LSTM and inaudible signals, the estimated AoA yields 11.33 dB SiSNR. Using the ground truth AoA achieves 11.52 dB SiSNR. Using location embeddings improves SiSNR to 12.74 dB, which is 1.22 dB higher than using the ground truth AoA. This is because the ground truth AoA only provides information about the direct path, but multipath information is also useful for beamforming. Further incorporating multi-task learning increases the SiSNR to 13.61 dB by jointly optimizing AoA and source separation. These results demonstrate each component in our system is useful and leveraging them all provides the best performance.

\para{Impact of sensing using inaudible signals:} Figure ~\ref{fig:aoa_err} shows that using inaudible signals together with audible signals decreases the AoA error by $1.8^{\circ}$ and $1.2^{\circ}$ over audible signal based sensing without mask and with mask, respectively. The reduced AoA error also translates into improved separation performance. 
Table~\ref{tab:noisy} shows that SiSNR decreases 0.6dB, 0.83dB and 0.54db without inaudible information for three different noise and interference setups. While inaudible signal based sensing is useful, it alone (denoted as \sysname (w/o audible)) performs less well. These results confirm that combining audible and inaudible signals for sensing yields the best performance. 

\subsubsection{Overall Performance}

In this section, we compare the overall performance of \sysname with several existing source separation methods by varying the background interference, users, SNR, and environments. 

\para{Vary interference:} Following UltraSE~\cite{UltraSE}, we compare our algorithm with Conv-TasNet and PHASEN under different numbers of interfering speakers and noise.  All schemes, are trained using the same data. Note that \sysname and PHASEN only require the target signal for training, whereas Conv-TasNet requires both the target signal and interference for training. To support multiple interferers, it takes the total interference from all interferers as the ground truth output for training. 
Table~\ref{tab:noisy} summarizes the performance in terms of SiSNR, PESQ and STOI. As it shows, our algorithm improves over Conv-TasNet and PHASENby 5.00 dB, and 1.39 dB, respectively, under only ambient noise. The corresponding numbers become 3.39 dB, 9.58 dB, and 9.69 dB, respectively, under 1 interfering speakers with ambient noise; and become 5.01 dB, 8.10 dB, and 9.02 dB, respectively, under 2 or more interfering speakers and ambient noise. The larger improvement over the existing approaches under interference is owing to the spatial embeddings learned from both audible and inaudible signals and multi-task learning. Conv-TasNet cannot perform well with only noise or more interference. PHASEN can deal with phase distortion caused by ambient noise, but cannot handle interference well.
\sysname can outperform all of them even in their target scenarios. \sysname also achieves better PESQ and STOI as it reduces the phase distortion.


\begin{table}[h!]
\setlength{\tabcolsep}{2pt}
\centering
\vspace*{-0.1in}
\begin{tabular}{ c| c | c | c | c}
    \toprule
    Environment & Model & SiSNR & PESQ & STOI \\
    \hline
    \multirow{5}{5em}{noise} & \sysname & \textbf{10.71} & \textbf{2.21} & \textbf{0.76} \\
    & SAMS(w/ GT AoA)  & 9.51 & 2.09 & 0.71 \\
    & SAMS(w/o inaudible) & 10.11 & 2.15 & 0.74 \\
    & SAMS(w/o audible) & 6.32 & 1.83 & 0.74 \\
    & Conv-TasNet & 5.71 & 1.76 & 0.60\\
    & PHASEN & 9.32 & 2.09 & 0.70\\
    \hline
    \multirow{5}{4em}{1 interferer + noise }& \sysname & \textbf{13.61} & \textbf{2.68} & \textbf{0.84}\\
    & SAMS(w/ GT AoA) & 11.52 & 2.49 & 0.79 \\
    & SAMS(w/o inaudible) & 12.78 & 2.54 & 0.81 \\
    & SAMS(w/o audible) & 9.93 & 2.14 & 0.77 \\
    & Conv-TasNet & 10.22 & 2.19 & 0.76\\
    & PHASEN & 4.03  & 1.60 & 0.53 \\
    \hline
    \multirow{5}{4em}{2 or 3 interferers + noise} & \sysname & \textbf{12.21} & \textbf{2.44} & \textbf{0.78}\\
    & SAMS(w/ GT AoA) & 10.33 & 2.38 & 0.75 \\
    & SAMS(w/o inaudible) & 11.67 & 2.41 & 0.77 \\
    & SAMS(w/o audible) & 7.01 & 1.92 & 0.62 \\
    & Conv-TasNet & 7.20 & 1.96 & 0.65\\
    & PHASEN & 4.11 & 1.69 & 0.53 \\
    \hline
\end{tabular}
\caption{Performance across various interference and noise scenarios}
\label{tab:noisy}
\vspace*{-0.4in}
\end{table}


\para{Vary gender pairs:} Next, we consider the impact of different sets of users speaking at the same time. Conceptually, female and male voices are different and have distinct resonant frequencies, so they are easier to separate out. This is confirmed in our evaluation. As shown in Figure~\ref{fig:user_SiSNR}, the SiSNR of separating from the female-male(F-M) pair is 14.21 dB. The SiSNR of separating from the female-female(F-F) and male-male(M-M) is lower, but still quite high: 13.37 dB and 12.67 dB, respectively. These results show that even when the interference is similar to the target signals, \sysname can still separate out the signals by taking advantage of location embeddings. 




\para{Vary SNR:} Then we vary the SNR of the target user from -6dB to 6dB by scaling the target signal. Each subset of a specific SNR includes all linear combinations of interference speech and noise.  As shown in Figure~\ref{fig:snr_SiSNR}, \sysname outperforms Conv-TasNet and PHASEN in all SNR scenarios by about 3dB and 5dB, respectively.  Even for the low SNR case, \sysname can separate the weaker target speech and improve SISNR to 7.09dB, which is sufficient for good audio quality in an online meeting.  


\para{Vary environments:} The target user speeches are collected from 5 different environments: living room, study room, cubicle, conference room and corridor.  Different environments only have different multipath, and can affect the performance of both AoA estimation and source separation. We train our DNN using the data collected from all environments except the one used for testing. Figure~\ref{fig:room} plots  the source separation performance for the new environment that is not in the training. As it shows, \sysname generalizes well to the new environment and its performance is fairly stable across different environments.


\para{Computation cost:} We run inference on three platforms to quantify the computation cost: a desktop GPU (NVIDIA GTX 2080 Ti), a desktop CPU (Intel Core I7-8700K), and a laptop CPU (Intel Core I5-5257U). These platforms represent common devices for online meetings. Processing 4-second audio takes 61ms on the desktop with GPU, 0.41s on the desktop with CPU, and 1.31s on the laptop with CPU using Pytorch without any optimization. With optimization through caching intermediate results and onnxruntime, our system can process 90-ms audio within 42 ms on the laptop with only CPU. 
The total latency is 132ms, which is smaller than 150 ms (\ie, the target latency requirement for VoIP), hence achieving real-time processing. 

%% file: conclusion.tex
\section{Conclusion}
\label{sec:conclusion}

We develop a novel system to combat acoustic interference for online meetings. It advances state-of-the-art in acoustic-based tracking by leveraging both audible and inaudible signals. Moreover, it uses multi-task learning to jointly estimate the AoA and separate the source. Our evaluation shows that our system significantly improves over the state-of-the-art. We believe our work is an important step towards enabling online meetings and classes under interference and noise, which have already been playing a major role in our daily lives. Moving forward, we are interested in exploring other context information to further improve the performance of online meetings.  

